\documentclass[amsmath,amssymb]{revtex4}

\usepackage{graphicx}
\usepackage{dcolumn}
\usepackage{bm}

\newcommand{\dis}{\displaystyle}
\newcommand{\bs}[1]{\mbox{\boldmath $#1$}}

\begin{document}

\title{Metropolis Algorithms in Generalized Ensemble}

\author{Yuko Okamoto}
  
\address{
~~\\
Department of Theoretical Studies\\
 Institute for Molecular Science\\
 Okazaki, Aichi 444-8585, Japan\\
 {\rm and}\\
 Department of Functional Molecular Science\\
 The Graduate University for Advanced Studies\\
 Okazaki, Aichi 444-8585, Japan}



\begin{abstract}
In complex systems such as spin systems and protein systems,
conventional
simulations in the canonical ensemble will get trapped in states
of energy local minima.  We employ the generalized-ensemble
algorithms in order to overcome this multiple-minima problem.  
Two
well-known generalized-ensemble algorithms, namely, multicanonical
algorithm and replica-exchange method, are
described.  We then present four new generalized-ensemble
algorithms as further extensions of the two methods.
Effectiveness of the new methods are illustrated with a Potts model,
Lennard-Jones fluid system, and protein system.
\end{abstract}

\maketitle


\section{Introduction}
In complex systems such as spin systems and protein systems,
conventional
simulations in the canonical ensemble will get trapped in states
of energy local minima at low temperatures.  We employ 
the {\it generalized-ensemble
algorithms} in order to overcome this multiple-minima problem  
(for reviews, see Refs.~\cite{RevHO}--\cite{RevSO}).
In a generalized-ensemble simulation,
each state is weighted by a non-Boltzmann probability
weight factor so that
a random walk in potential energy space may be realized.
The random walk allows the simulation to escape from any
energy barrier and to sample much wider configurational space than
by conventional methods.
Monitoring the energy in a single simulation run, one can
obtain not only
the global-minimum-energy state but also canonical ensemble
averages as functions of temperature by the single-histogram \cite{FS1}
and multiple-histogram \cite{FSWHAM} reweighting techniques.

One of the most well-known generalized-ensemble methods is perhaps
{\it multicanonical algorithm} (MUCA) \cite{MUCA}.
(The method is also referred to as {\it entropic sampling}
\cite{Lee}, {\it adaptive umbrella sampling} \cite{MZ} 
{\it of the potential energy} \cite{BK},
{\it random walk algorithm} \cite{Landau},
and {\it density of states Monte Carlo} \cite{dePablo}.)
MUCA was first introduced to the molecular
simulation field in Ref.~\cite{HO}.
Since then MUCA has been extensively
used in many applications in
protein and related systems (for a review, see, e.g.,
Ref.~\cite{RevHO}).


The {\it replica-exchange method} (REM) \cite{RE1,RE2}
is another widely used generalized-ensemble algorithm.
(Closely related methods were
independently developed in Refs.~\cite{RE3}--\cite{JWK}.
REM is also referred to as
{\it multiple Markov chain method} \cite{Whit}
and {\it parallel tempering} \cite{STrev}.
For recent reviews with detailed references about the method, see,
e.g., Refs.~\cite{RevMSO,IBArev}.)
REM has also been introduced to protein systems
\cite{H97}--\cite{Berne}.


Both MUCA and REM are already very powerful, but we have also
developed several new generalized-ensemble algorithms
as further extensions of MUCA and/or REM \cite{SKO}--\cite{BNO03}.

In this article, 
we first describe the two familiar methods:
MUCA and REM.  We then present some of our new generalized-ensemble algorithms.
The effectiveness of these methods is illustrated with
a 2-dimensional Potts model, Lennard-Jones fluid system, and protein system.

\section{Methods}
In the regular canonical ensemble with a given
inverse temperature
$\beta \equiv 1/k_BT$ ($k_B$ is the Boltzmann
constant),
the probability distribution of potential
energy $E$ is given by
\begin{equation}
P_B(E;T)\ ~\propto ~n(E)~ W_B (E;T)~
\equiv \ n(E)~ e^{- \beta E}~,
\label{pb}
\end{equation}
where $n(E)$ is the density of states.
Since the density of states $n(E)$ is a rapidly increasing function of $E$
and the Boltzmann factor $W_B (E;T)$ decreases exponentially with $E$,
the probability distribution $P_B(E;T)$ has a bell-like shape in general.
A Monte Carlo (MC) simulation based on the Metropolis 
algorithm \cite{Metro} generates states in the canonical
ensemble with the following transition probability  
from a state $x$
with energy $E$ to a state $x^{\prime}$ with energy $E^{\prime}$:
\begin{equation}
w(x \rightarrow x^{\prime})
= {\rm min} \left(1,\frac{W_{B}(E^{\prime};T)}{W_{B}(E;T)}\right)
= {\rm min} \left(1,e^{-\beta (E^{\prime} - E)}\right)~.
\label{Eqn3}
\end{equation}

However, it is very difficult to obtain
canonical
distributions at low temperatures with this conventional
Metropolis algorithm.
This is because the thermal fluctuations at
low temperatures are small and the
simulation will certainly
get trapped in states of energy local minima.

In the ``multicanonical ensemble'' \cite{MUCA}, on the other hand, 
the probability distribution of potential
energy is {\it defined} as follows so that a uniform flat
distribution of $E$ may be obtained:
\begin{equation}
P_{mu} (E) ~\propto ~ n (E)~ W_{mu} (E) \equiv {\rm constant}~.
\label{pmu}
\end{equation}
Hence, the multicanonical weight factor
$W_{mu} (E)$ is inversely proportional to the density of
states, and the Metropolis criterion for 
the multicanonical MC simulations is based on
the following transition probability:
\begin{equation}
w(x \rightarrow x^{\prime})
= {\rm min} \left(1,\frac{W_{mu}(E^{\prime})}{W_{mu}(E)}\right)
= {\rm min} \left(1,\frac{n(E)}{n(E^{\prime})}\right)~.
\label{Eqn3b}
\end{equation}
Because the MUCA weight factor $W_{mu} (E)$
is not {\it a priori} known, however, one has to
determine it for each system by iterations of trial
simulations.

After the optimal MUCA weight factor is obtained, one
performs a long MUCA simulation once.  By monitoring
the potential energy throughout the simulation, one can find the
global-minimum-energy state.
Moreover, by using the obtained histogram $N_{\rm mu}(E)$ of the
potential energy distribution $P_{\rm mu}(E)$,
the expectation value of a physical quantity $A$
at any temperature $T=1/k_{\rm B} \beta$
can be calculated from
\begin{equation}
<A>_{T} \ = \frac{\dis{\sum_{E}~A(E)~n(E)~e^{-\beta E}}}
{\dis{\sum_{E} ~n(E)~e^{-\beta E}}}~,
\label{eqn18}
\end{equation}
where the best estimate of the
density of states is given by the single-histogram
reweighting techniques (see Eq.~(\ref{pmu})) \cite{FS1}:
\begin{equation}
 n(E) = \frac{N_{\rm mu}(E)}{W_{\rm mu}(E)}~.
\label{eqn17}
\end{equation}

The system for {\it replica-exchange method} (REM) \cite{RE1,RE2}
consists
of $M$ non-interacting copies, or replicas, of the original
system in canonical ensemble at $M$ different
temperatures $T_m$ ($m=1, \cdots, M$).
We arrange the replicas so that there is always one replica at each temperature.
Then there is a one-to-one correspondence between replicas and temperatures.
Let $X = \left\{ \cdots, x_m^{[i]},\cdots \right\}$ stand for a state 
in this generalized ensemble.
Here, the superscript $i$ and the subscript $m$ in $x_m^{[i]}$ label 
the replica and the temperature, respectively.
A simulation of REM is then realized by alternately performing the 
following two steps. 
Step 1: Each replica in the canonical ensemble at a fixed temperature 
is simulated simultaneously and independently for a certain number 
of MC 
steps.
Step 2: A pair of replicas, say $i$ and $j$, which are at neighboring 
temperatures, say $T_m$ and $T_{m+1}$, respectively,
are exchanged: 
$X = \left\{\cdots, x_m^{[i]}, \cdots, x_{m+1}^{[j]}, \cdots \right\} 
\to X^\prime = \left\{\cdots, x_m^{[j]}, \cdots, x_{m+1}^{[i]}, 
\cdots \right\}$.
The transition probability of this replica exchange
is given by the following Metropolis criterion:
\begin{equation}
w(X \rightarrow X^{\prime})
= {\rm min} (1,e^{- \Delta})~,
\label{eqn1}
\end{equation}
where
\begin{equation}
\Delta \equiv \left(\beta_{m+1} - \beta_m \right)
              \left(E\left(q^{[i]}\right)
                  - E\left(q^{[j]}\right)\right)~.
\label{eqn14}
\end{equation}
  
From the results of a long REM production run, one can obtain
the canonical ensemble average of a physical quantity $A$ as a function
of temperature from Eq.~(\ref{eqn18}), where the
density of states is given by
the multiple-histogram reweighting techniques \cite{FSWHAM}
as follows.
Let $N_m(E)$ and $n_m$ be respectively
the potential-energy histogram and the total number of
samples obtained at temperature $T_m=1/k_{\rm B} \beta_m$
($m=1, \cdots, M$).
The best estimate of the density of states is then
given by \cite{FSWHAM}
\begin{equation}
n(E) = \frac{\dis{\sum_{m=1}^M ~g_m^{-1}~N_m(E)}}
{\dis{\sum_{m=1}^M ~g_m^{-1}~n_m~e^{f_m-\beta_m E}}}~,
\label{Eqn8a}
\end{equation}
where
\begin{equation}
e^{-f_m} = \sum_{E} ~n(E)~e^{-\beta_m E}~.
\label{Eqn8b}
\end{equation}
Here, $g_m = 1 + 2 \tau_m$,
and $\tau_m$ is the integrated
autocorrelation time at temperature $T_m$.
Note that
Eqs.~(\ref{Eqn8a}) and
(\ref{Eqn8b}) are solved self-consistently
by iteration \cite{FSWHAM} to obtain
the dimensionless Helmholtz free energy $f_m$
and the density of states $n(E)$.

We now introduce new generalized-ensemble algorithms
that combine the merits of MUCA and REM.
In the {\it replica-exchange multicanonical algorithm}
(REMUCA) \cite{SO3,MSO03} 
%
we first perform a short REM
simulation (with $M$ replicas)
to determine the MUCA
weight factor and then perform with this weight
factor a regular MUCA simulation with high statistics.
The first step is accomplished by the multiple-histogram reweighting
techniques \cite{FSWHAM}.
Let $N_m(E)$ and $n_m$ be respectively
the potential-energy histogram and the total number of
samples obtained at temperature $T_m=1/k_{\rm B} \beta_m$ of the REM run.
The density of states $n(E)$, or the inverse of the MUCA
weight factor, is then given by solving
Eqs.~(\ref{Eqn8a}) and (\ref{Eqn8b}) self-consistently by iteration
\cite{FSWHAM}.
The formulation of REMUCA is simple and straightforward, but
the numerical improvement is great, because the weight factor
determination for MUCA becomes very difficult
by the usual iterative processes for complex systems.

While multicanonical simulations are
usually based on local
updates, a replica-exchange process can be considered to be a
global update, and global updates enhance the sampling further.
Here, we present a further modification of REMUCA and refer to the
new method as {\it multicanonical replica-exchange method}
(MUCAREM) \cite{SO3,MSO03}.  In MUCAREM the final production run is not a
regular multicanonical simulation but a replica-exchange simulation
with a few replicas
in the multicanonical ensemble.
Because multicanonical simulations cover much wider energy
ranges than regular canonical simulations, the number of
required replicas for the production run of MUCAREM is
much less than that for the regular REM,
and we can keep the merits of
REMUCA (and improve the sampling further).
The details of REMUCA and MUCAREM can be found in Ref.~\cite{RevMSO}

Besides canonical ensemble, MC simulations in isobaric-isothermal 
ensemble \cite{mcd72} 
are also extensively used. 
This is 
because most experiments are carried out 
under the constant pressure and constant temperature conditions. 
The distribution ${\rm P}_{NPT}(E,V)$ 
for $E$ and $V$ is given by 
\begin{equation}
 {\rm P}_{NPT}(E,V) = n(E,V) {\rm e}^{-\beta_0 H}~. 
\label{eqn2}
\end{equation}
Here, the density of states $n(E,V)$ is given
as a function of both $E$ and $V$, 
and $H$ is the ``enthalpy'':
\begin{equation}
 H = E+P_0V~, 
\label{eqn3}
\end{equation}
where $P_0$ is the pressure at which simulations are performed. 
This ensemble has 
bell-shaped distributions in both $E$ and $V$. 

We now introduce the idea of the multicanonical technique 
into the isobaric-isothermal ensemble MC method and 
refer to this generalized-ensemble algorithm as 
the {\it multibaric-multithermal algorithm} \cite{OO03}. 
This MC simulation performs random walks 
in volume space as well as in potential energy space. 
   
In the multibaric-multithermal ensemble, 
each state is sampled by a weight factor 
$W_{\rm mbt}(E,V)\equiv \exp \{-\beta_0 H_{\rm mbt}(E,V)\}$ 
($H_{\rm mbt}$ is referred to as the multibaric-multithermal enthalpy) 
so that a uniform distribution in both potential energy 
and volume 
is obtained: 
\begin{equation}
 {\rm P_{mbt}}(E,V) = n(E,V) W_{\rm mbt}(E,V) = {\rm constant}~. 
\label{eqn5}
\end{equation}
We call $W_{\rm mbt}(E,V)$ the multibaric-multithermal weight factor. 

In order to perform the multibaric-multithermal MC simulation, 
we follow 
the conventional isobaric-isothermal MC techniques \cite{mcd72}. 
In this method, we perform Metropolis sampling on the scaled coordinates 
${\bs s}_i = L^{-1} {\bs r}_i$ 
(${\bs r}_i$ are the real coordinates) 
and the volume $V$ (here, the particles are placed in a cubic box of
a side of size $L \equiv \sqrt[3]{V}$). 
The trial moves of the scaled coordinates 
from ${\bs s}_i$ to ${\bs s'}_i$ and 
of the volume from $V$ to $V'$ are generated by uniform random numbers. 
The enthalpy is accordingly changed 
from $H(E({\bs s}^{(N)},V),V)$ 
to $H'(E({\bs s}'^{(N)},V'),V')$ 
by these trial moves. 
The trial moves will be accepted with the probability 
\begin{equation}
w(x \rightarrow x^{\prime})
= {\rm min} 
\left(1,\exp[-\beta_0 \{ 
 H' - H - N k_B T_0 \ln(V'/V)
 \}]\right)~, 
 \label{accibt:eq}
\end{equation}
where $N$ is the total number of particles in the system.

Replacing $H$ by $H_{\rm mbt}$, 
we can perform the multibaric-multithermal MC simulation. 
The trial moves of ${\bs s}_i$ and $V$ are generated in the same way 
as in the isobaric-isothermal MC simulation. 
The multibaric-multithermal enthalpy is changed 
from $H_{\rm mbt}(E({\bs s}^{(N)},V),V)$ 
to $H'_{\rm mbt}(E({\bs s}'^{(N)},V'),V')$ 
by these trial moves. 
The trial moves will now be accepted with the probability 
\begin{equation}
w(x \rightarrow x^{\prime})
= {\rm min} 
\left(1,\exp[-\beta_0 \{ 
 H'_{\rm mbt} - H_{\rm mbt} - N k_B T_0 \ln(V'/V)
 \}]\right)~. 
 \label{acc:eq}
\end{equation}

While MUCA yields a flat distribution in potential energy
and performs a random walk in potential energy space,
we can, in principle, choose any other variable and 
induce a random walk in that variable.
One such example is the {\it multi-overlap algorithm}
\cite{BNO03}.  Here, we choose a protein system and
define the overlap in 
the space of dihedral angles by, as it was already used 
in~\cite{HMO97}, 
\begin{equation} \label{dihedral_ovl}
 q\ =\ (n - d)/n\ ,
\end{equation}
where $n$ is the number of dihedral angles and $d$ is the distance 
between configurations defined by
\begin{equation} \label{dihedral_d}
  d\ =\ ||v-v^1||\ =\ {1\over \pi} \sum_{i=1}^n d_a(v_i,v_i^1)\ .
\end{equation}
Here, $v_i$ is our generic notation for the dihedral angle $i$,
$-\pi < v_i \le \pi$, and $v^1$ is the vector of dihedral angles
of the reference configuration. 
The distance $d_a(v_i,v_i')$ between two angles is defined by
\begin{equation} \label{d_a}
  d_a(v_i,v_i')\ =\ \min (|v_i-v_i'|,2\pi-|v_i-v_i'|)\ .
\end{equation}

We want to simulate the system with 
weight factors that lead to a flat distribution in
the dihedral distance $d$, and hence to a random walk process in 
$d$: 
\begin{equation} \label{RWC}
d < d_{\min}\ \to\ d> d_{\max} ~~{\rm and~~back}\ . 
\end{equation}
Here, $d_{\min}$ is chosen sufficiently small so that one can
claim that the reference configuration has been reached.
The value of $d_{\max}$ has to be sufficiently large 
to introduce a considerable amount of disorder.

Moreover, we can define a weight factor that
leads to a random-walk process between two configurations. 
This multi-overlap simulation
allows a detailed study of the
transition states between the two configurations,
whereas a random walk in
energy space of a regular MUCA simulation
may miss the transition state (see Ref.~\cite{BNO03} for 
details).

\begin{figure}
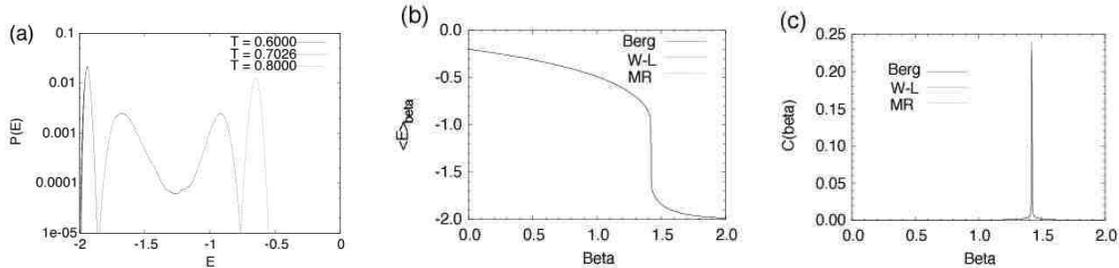

\includegraphics[width=5.0cm,keepaspectratio]{fg1.epsf}
\includegraphics[width=5.0cm,keepaspectratio]{e.epsf}
\includegraphics[width=5.0cm,keepaspectratio]{c.epsf}
\caption{(a) Probability distributions of energy of 2-dimensional
10-state Potts model at three temperatures: $T=0.6000$,
0.7026, and 0.8000, and (b) average energy and (c) specific
heat as functions of inverse temperature $\beta = 1/T$. 
The results were obtained from a multicanonical MC simulation.
For (b) and (c), the results from three methods of multicanonical
weight factor determination are superimposed.
Berg stands for Berg's method \cite{MUCAW}, 
W-L for Wang-Landau's method \cite{Landau} and MR for
MUCAREM \cite{SO3}}
\label{fig1}
\end{figure}


  
\begin{figure}
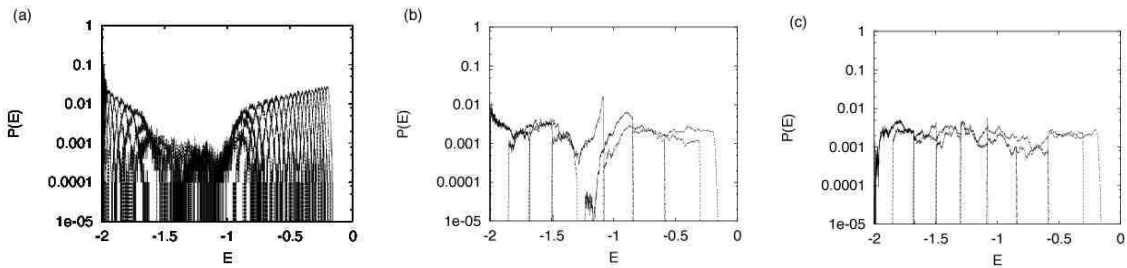

\includegraphics[width=5.0cm,keepaspectratio]{fg3a.epsf}
\includegraphics[width=5.0cm,keepaspectratio]{fg3b.epsf}
\includegraphics[width=5.0cm,keepaspectratio]{fg3d.epsf}
\caption{Probability distributions of energy
for the 2-dimensional 10-state Potts model:
(a) the results of REM simulation with 32 replicas and
(b) and (c) iterations of MUCAREM simulations with 8 replicas.}
\label{fig2}
\end{figure}

\section{Results}
We now present the results of our simulations
based on
the algorithms described in the previous section.

\begin{figure}
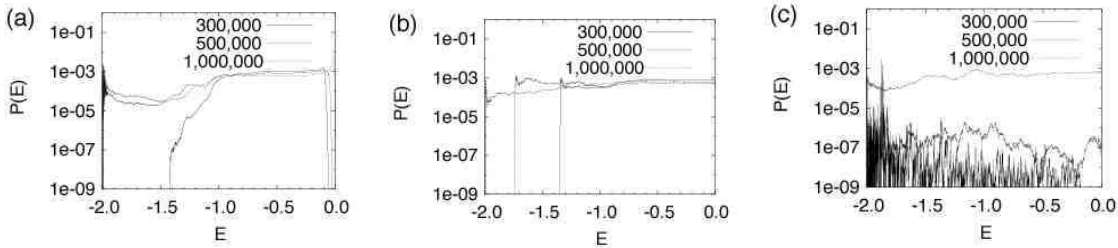

\includegraphics[width=5.0cm,keepaspectratio]{mr.hst.epsf}
\includegraphics[width=5.0cm,keepaspectratio]{berg.hst.epsf}
\includegraphics[width=5.0cm,keepaspectratio]{wl.hst.epsf}
\caption{Probability distribution of energy during the iterative 
process of multicanonical weight factor determination
for the 2-dimensional
10-state Potts model.
The results after 300,000 MC sweeps, 500,000 MC sweeps,
and 1,000,000 MC sweeps are superimposed.
(a) MUCAREM \cite{SO3}, (b) Berg's method \cite{MUCAW}, 
and (c) Wang-Landau's method \cite{Landau}.}
\label{fig3}
\end{figure}

The first example is a spin system.
We studied the 2-dimensional 10-state Potts model
\cite{NSMO}.
The lattice size was $34 \times 34$.
This system exhibits a first-order phase transition
\cite{Bax}.
In Fig.~\ref{fig1} we show the probability distributions
of energy at three tempeartures (above the critical
temperature $T_C$, at $T_C$, and
below $T_C$) and average energy and specific heat as functions
of inverse temperature.  
All these results imply
that the system indeed undergoes a first-order phase transition.


Iterations of MUCAREM (and REMUCA) can be used to obtain
an optimal MUCA weight factor \cite{MSO03}.
In Fig.~\ref{fig2} we show the results of our MUCA
weight factor determination by MUCAREM.  We first made a REM
simulation of 10,000 MC sweeps (for each replica)
with 32 replicas (Fig.~\ref{fig2}(a)).
Using the obtained energy
distributions, we determined the (preliminary)
MUCA weight factor, or the density of states, by the
multiple-histogram reweighting techniques of 
Eqs.~(\ref{Eqn8a}) and (\ref{Eqn8b}).
Because the trials of
replica exchange are not accepted near the critical
temperature for first-order phase transitions, the probability
distributions in Fig.~\ref{fig2}(a) for the energy range from
$\sim - 1.5$ to $\sim -1.0$ fails to have sufficient overlap,
which is required for successful application of REM.
This means that the MUCA weight factor
in this energy range thus determined is of ``poor quality.''
With this MUCA weight factor, however, we made iterations of
three MUCAREM simulations of 10,000 MC sweeps (for each
replica) with 8 replicas (Fig.~\ref{fig2}(b) for the first
iteration and Fig.~\ref{fig2}(c) for the third iteration).
In Fig.~\ref{fig2}(b) we see that the distributions are not
completely flat, reflecting the poor quality in the
phase-transition region.  This problem is rapidly
rectified as iterations continue, and the distributions
are completely flat in Fig.~\ref{fig2}(c), which gives an
optimal MUCA weight factor in the entire energy range by
the multiple-histogram reweighting techniques.

Besides MUCAREM the methods of Berg \cite{MUCAW}
and Wang-Landau \cite{Landau} are also effective for the
determination of the MUCA weight factor (or the density of states).
In Fig.~\ref{fig3} we compare how fast these three methods
converge to yield an optimal MUCA weight factor, or a flat
distribution.
Each figure shows three curves superimposed that correspond
to the (hot-start) MUCA simulation with the weight factor that was
``frozen'' after
300,000 MC seeps, 500,000 MC sweeps, and 1,000,000 MC sweeps
of iterations of the weight factor determination.
While the results of MUCAREM and Berg's method are similar, 
those of
Wang-Landau method have quite different behavior.
After 300,000 MC sweeps, MUCAREM and Berg's method gives
reasonably flat distribution for  $E>-1.0$ and $E>-1.3$, 
respectively, whereas
Wang-Landau method gives a distribution that is quite spiky
(but already covers low-energy regions).
After 500,000 MC sweeps, MUCAREM essentially gives a flat
distribution in the entire energy range and Berg's method
for $E>-1.7$, while Wang-Landau method
also covers the entire range (though still very rugged).
After 1,000,000 MC sweeps, the three methods all give
reasonably flat distributions.
Details will be published elsewhere \cite{NSMO}.

%
We now present the results of our multibaric-multithermal
simulation.
We considered a Lennard-Jones 12-6 potential system. 
We used 500 particles ($N=500$)
in a cubic unit cell with periodic boundary conditions. 
The length and the energy are scaled in units of 
the Lennard-Jones diameter $\sigma$ and 
the minimum value of the potential $\epsilon$, respectively. 
We use an asterisk ($*$) for the reduced quantities 
such as 
the reduced length $r^* = r/\sigma$, 
the reduced temperature $T^*=k_{\rm B}T/\epsilon$, 
the reduced pressure $P^*=P\sigma^3/\epsilon$, and 
the reduced number density $\rho^*=\rho \sigma^3$ ($\rho \equiv N/V$). 

We started the iterations of the
multibaric-multithermal weight factor determination
from a regular isobaric-isothermal
simulation
at $T_0^* = 2.0$ and $P_0^* = 3.0$. 
In one MC sweep we made the trial moves of all particle coordinates and the volume 
($N+1$ trial moves altogether).  For each trial move the Metropolis evaluation
of Eq. (\ref{acc:eq}) was made.
Each iteration of the weight factor determination consisted 
of 100,000 MC sweeps.
In the present case, it was required to make 12 iterations to
get an optimal weight factor
$W_{\rm mbt}(E,V)$.
We then performed a long multibaric-multithermal MC simulaton 
of 400,000 MC sweeps 
with this $W_{\rm mbt}(E,V)$. 


  
\begin{figure}
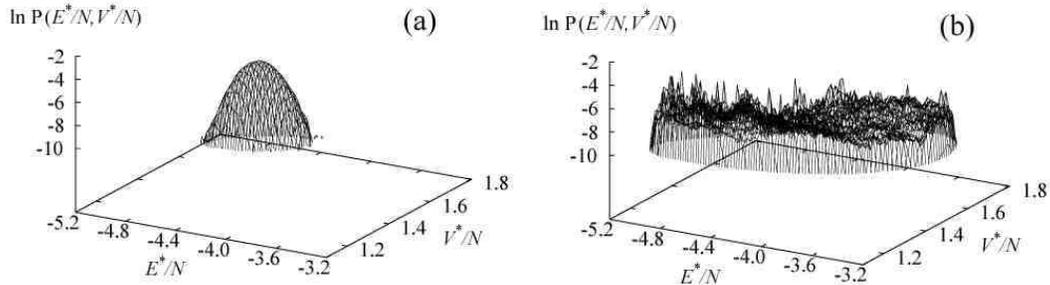

\includegraphics[width=7.0cm,keepaspectratio]{fig1a.epsf}
\includegraphics[width=7.0cm,keepaspectratio]{fig1b.epsf}
\caption{
(a) The probability distribution ${\rm P}_{NPT}(E^*/N,V^*/N)$ 
in the isobaric-isothermal simulation 
at $(T^*,P^*)=(T_0^*,P_0^*)=(2.0,3.0)$ and
(b) the probability distribution ${\rm P_{\rm mbt}}(E^*/N,V^*/N)$ 
in the multibaric-multithermal simulation. 
}
\label{dis:fig}
\end{figure}
   
\begin{figure}
\includegraphics[width=14.0cm,keepaspectratio]{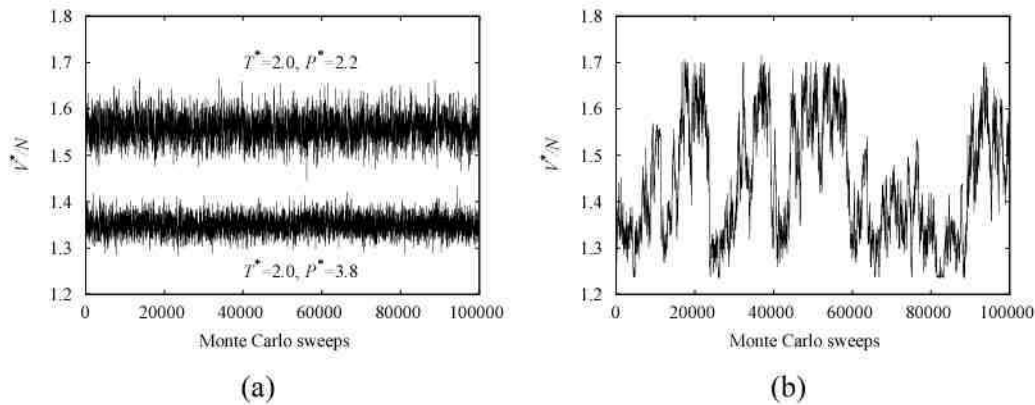}
\caption{
The time series of $V^*/N$ from 
(a) the conventional isobaric-isothermal MC simulations 
at $(T^*,P^*)=(2.0,2.2)$ and at $(T^*,P^*)=(2.0,3.8)$ and 
(b) the multibaric-multithermal MC simulation. 
}
\label{vol:fig}
\end{figure}

\begin{figure}
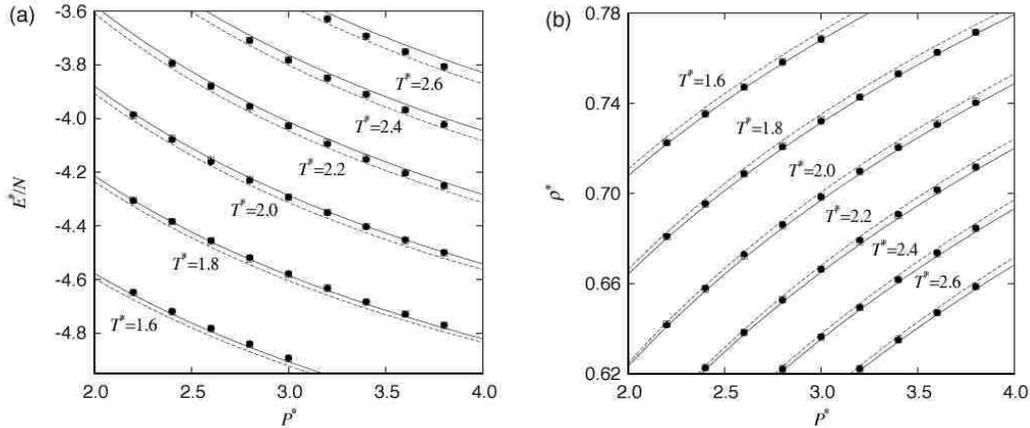

\includegraphics[width=7.0cm,keepaspectratio]{fig4.epsf}
\includegraphics[width=7.0cm,keepaspectratio]{fig5.epsf}
\caption{
(a) Average potential energy per particle $<E^*/N>_{NPT}$ 
and (b) average density $<\rho^*>_{NPT}$ 
at various temperature and pressure values. 
Filled circles: Multibaric-multithermal MC simulations. 
Open squares: Conventional isobaric-isothermal MC simulations. 
Solid line: Equation of states calculated by Johnson et al. \cite{jzg93}. 
Broken line: Equation of states calculated by Sun and Teja \cite{sj93}. 
}
\label{rwted:fig}
\end{figure}

Figure \ref{dis:fig} shows the probability distributions of 
$E^*/N$ and $V^*/N$. 
Figure \ref{dis:fig}(a) is the probability 
distribution ${\rm P}_{NPT}(E^*/N,V^*/N)$ 
from the isobaric-isothermal simulation first carried out in the
process
(i.e., $T_0^*=2.0$ and $P_0^*=3.0$). 
It is a bell-shaped distribution. 
On the other hand, Fig. \ref{dis:fig}(b) is the probability 
distribution ${\rm P_{\rm mbt}}(E^*/N,V^*/N)$ 
from the multibaric-multithermal simulation finally performed. 
It shows a flat distribution, and  
the multibaric-multithermal MC simulation indeed 
sampled the configurational space 
in wider ranges of energy and volume 
than the conventional isobaric-isothermal MC simulation. 


Figure \ref{vol:fig} shows 
the time series of $V^*/N$. 
In Fig. \ref{vol:fig}(a) we show the results of the conventional
isobaric-isothermal simulations at
$(T^*,P^*)=(2.0,2.2)$ and (2.0,3.8), while in
Figure \ref{vol:fig}(b) we give those of the multibaric-multithermal
simulation.
The volume fluctuations 
in the conventional isobaric-isothermal MC simulations 
are only in the range of 
$V^*/N = 1.3 \sim 1.4$ and $V^*/N = 1.5 \sim 1.6$ 
at $P^*=3.8$ and at $P^*=2.2$, respectively. 
On the other hand, 
the multibaric-multithermal MC simulation performs a random walk 
that covers even a wider volume range.

We calculated the ensemble averages of 
potential energy per particle, $<E^*/N>_{NPT}$, and
density, $<\rho^*>_{NPT}$, at various temperature and pressure values
by the reweighting techniques.
They are shown in Fig. \ref{rwted:fig}.
The agreement between the multibaric-multithermal data and 
isobaric-isothermal data are excellent 
in both $<E^*/N>_{NPT}$ and $<\rho^*>_{NPT}$. 

The important point is that 
we can obtain any desired isobaric-isothermal distribution 
in wide temperature and 
pressure ranges ($T^*=1.6 \sim 2.6$, $P^*=2.2 \sim 3.8$) 
from a single simulation run by the multibaric-multithermal MC algorithm. 
This is an outstanding advantage over
the conventional isobaric-isothermal MC algorithm, 
in which simulations have to be carried out 
separately at each temperature and pressure, because the reweighting
techniques based on the isobaric-isothermal simulations
can give correct results only for
narrow ranges of temperature and pressure values.

\begin{figure}
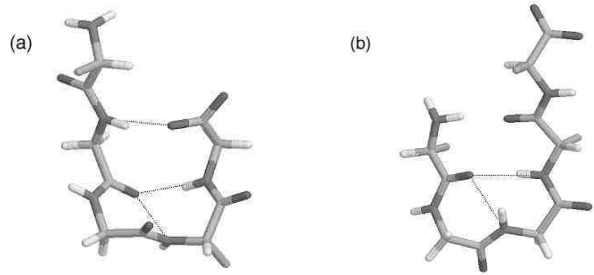

\includegraphics[width=4.0cm,keepaspectratio]{A4_bw.epsf}
\includegraphics[width=4.0cm,keepaspectratio]{B4_bw.epsf}
\caption{(a) Reference configuration~1 and (b) reference
configuration~2.  Only backbone structures are shown.
The N-terminus is on the left-hand side and the C-terminus on the
right-hand side.  The dotted lines stand for hydrogen bonds.
The figures were created with RasMol \cite{RasMol}.}
\label{fig7}
\end{figure}

\begin{figure}
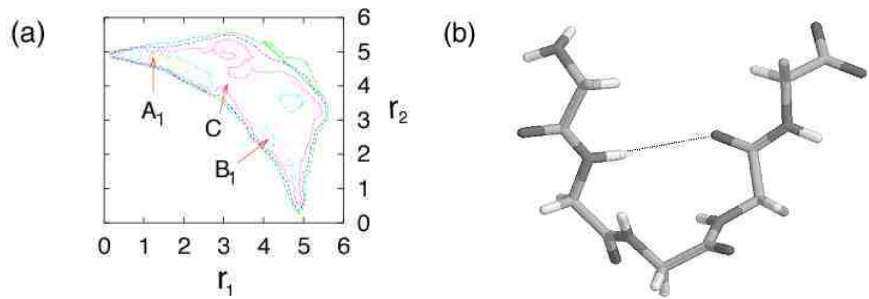

\includegraphics[height=0.18\textheight]{F_rms-2.epsf}
\includegraphics[height=0.17\textheight]{C_bw.epsf}
\caption{(a) Free-energy landscape of Met-enkephalin
at $T=250$ K with respect to rms
distances (\AA) from the two reference configurations,
$F(r_1,r_2)$.
The labels A$_1$ and B$_1$ indicate the positions for the local-minimum
states at $T=250$ K that originate from the reference configuration 1
and the reference configuration 2, respectively.  The label C 
stands for the saddle point that corresponds to
the transition state. 
(b) The transition state, C, between reference configurations
1 and 2.  See the caption of
Fig.~\ref{fig7} for details.}
\label{fig8}
\end{figure}

The third example is a system of a biopolymer.
A brain peptide Met-enkephalin has the amino-acid sequence Tyr-Gly-Gly-Phe-Met.
We fix the peptide-bond dihedral angles $\omega$ to $180^{\circ}$, which implies
that the total number of variable dihedral angles is $n=19$. We
neglect the solvent effects as in previous works. The low-energy
configurations of Met-enkephalin in the gas phase have been 
classified into several groups of similar structures~\cite{OKK92,MHO98}. Two
reference configurations, called configuration~1 and 
configuration~2, are used in the following and depicted in
Fig.~\ref{fig7}. 
Configuration~1 has a $\beta$-turn structure
with hydrogen bonds between Gly-2 and Met-5, and configuration~2 a
$\beta$-turn with a hydrogen bond between Tyr-1 and Phe-4~\cite{MHO98}. 
Configuration 1 corresponds to the global-minimum-energy state
and configuration 2 to the second lowest-energy state.
The distance between the two configurations is $d=6.62$ 
and the values of the potential energy (ECEPP/2 \cite{ECEP3}) for configuration 1 
and configuration 2 are $-10.72$ kcal/mol and $-8.42$ kcal/mol, respectively.

We analyze the free-energy landscape~\cite{HOO99} 
from the results of our multi-overlap simulation at 300 K
that performs a random walk between configurations 1 and 2.
We study the landscape with respect to some
reaction coordinates (and hence it should be called
the potential of mean force).
In order to study the transition states between reference
configurations 1 and 2, we first plotted the free-energy
landscape with respect to the dihedral
distances $d_1$ and $d_2$ of Eq.~(\ref{dihedral_d}).
However, we did not observe any transition saddle point.
A satisfactory
analysis of the saddle point becomes possible when
the root-mean-square (rms) distance (instead of the 
dihedral distance) is used.
Figure~\ref{fig8} shows contour lines of the free 
energy reweighted to $T=250$ K, which is close to
the folding temperature~(\cite{HMO97,BNO03}).
Here, the free energy $F(r_1,r_2)$ is defined by
\begin{equation} \label{FE}
F(r_1,r_2)=-k_B T \ln P(r_1,r_2)~,
\end{equation}
where $r_1$ and $r_2$ are the rms distances 
from the reference
configuration 1 and the reference configuration 2, respectively,
and $P(r_1,r_2)$ is the (reweighted) probability at $T=250$ K
to find the peptide with values $r_1,r_2$.
The probability was calculated from the two-dimensional
histogram of bin size 0.06 \AA $\times$ 0.06 \AA.
The contour lines were plotted every $2 k_B T$ ($=0.99$ kcal/mol
for $T=250$ K).

\begin{table}[!t]
\caption{Free energy, internal energy, entropy multiplied by temperature
at $T=250$ K (all in kcal/mol) at the two local-minimum states (A$_1$ and B$_1$)
and the transition state (C) in Fig.~\ref{fig8}(a).
The rms distances, $r_1$ and $r_2$, are in \AA.
\label{tab_FE} } 
\vspace{2mm}
\begin{tabular}{lrrr} 
\hline
Coordinate $(r_1,r_2)$ & $F$ & $U$ & $-TS$ \\ 
\hline
A$_1$ (1.23, 4.83) & 0 & ~~~$-5.4$ & ~~~~~~5.4 \\ \hline
B$_1$ (4.17, 2.43) & 1.0 & $-3.5$ & 4.5 \\ \hline
C (3.09, 4.05) & ~~~2.2 & $-0.8$ & 3.0 \\ \hline
\end{tabular}
\end{table}
   
Note that the reference configurations 1 and 2, which are respectively
located at $(r_1,r_2)=(0,4.95)$ and $(4.95,0)$, are not
local minima in free energy at the finite temperature ($T=250$ K) 
because of the entropy contributions.
The corresponding local-minimum states at A$_1$ and B$_1$
still have the characteristics of the reference configurations
in that they have backbone hydrogen bonds between Gly-2 and
Met-5 and between Tyr-1 and Phe-4, respectively.

The transition state C in Fig.~\ref{fig8}(a) should have
intermediate structure between configurations 1 and 2.
In Fig.~\ref{fig8}(b) we show a typical backbone structure of
this transition state.
We see the backbone hydrogen bond between Gly-2 and Phe-4.
This is precisely the expected intermediate structure between
configurations 1 and 2, because going from configuration 1 to
configuration 2 we can follow the backbone hydrogen-bond
rearrangements:  The hydrogen bond between
Gly-2 and Met-5 of configuration 1 is broken, Gly-2 forms a hydrogen 
bond with Phe-4 (the
transition state), this new hydrogen bond is broken, and finally
Phe-4 forms a hydrogen bond with Tyr-1 (configuration 2).

In Ref.~\cite{MHO98} the low-energy conformations of Met-enkephalin
were studied in detail and they were classified into several 
groups of similar structures based on the pattern of backbone
hydorgen bonds.  It was found there that below
$T=300$ K there are two dominant groups, which correspond to
configurations 1 and 2 in the present article.
Although much less conspicuous, the third most populated
structure is indeed the group that is identified to be the 
transition state in the present work.

In table~\ref{tab_FE} we list the numerical values of the
free energy, internal energy, and entropy multiplied by
temperature at the two local-minimum states (A$_1$ and B$_1$
in Fig.~\ref{fig8}(a)) and the transition state
(C in Fig.~\ref{fig8}(a)).  
Here, the internal energy $U$ is defined by the (reweighted) average
ECEPP/2 \cite{ECEP3} potential energy: 
\begin{equation} \label{U}
U(r_1,r_2) = <E(r_1,r_2)>_T~.
\end{equation}
The entropy $S$ was then calculated by
\begin{equation} \label{TS}
S(r_1,r_2) = \frac{1}{T} \left[U(r_1,r_2) - F(r_1,r_2)\right]~.
\end{equation}
The free energy was normalized
so that the value at A$_1$ is zero.  

The state A$_1$ can be considered to be ``deformed'' configuration 1,
and B$_1$ deformed configuration 2 due to the entropy effects, whereas
C is the transition state between A$_1$ and B$_1$.
Among these three points, the free energy $F$ and
the internal energy $U$ are the lowest at A$_1$, while the
entropy contribution $-TS$ is the lowest at C.
The free energy difference $\Delta F$, internal energy difference
$\Delta U$, and entropy contribution difference $-T \Delta S$
are 1.0 kcal/mol, 1.9 kcal/mol, and $-0.9$ kcal/mol between B$_1$ and A$_1$,
2.2 kcal/mol, 4.6 kcal/mol, and $-2.4$ kcal/mol between C and A$_1$,
and 1.2 kcal/mol, 2.7 kcal/mol, and $-1.5$ kcal/mol between C and B$_1$.
Hence, the internal energy contribution and the entropy contribution to
free energy are opposite in sign and the magnitude of the
former is roughly twice as that of the latter at this temperature.

\section{Conclusions}
In this article we have described the
formulations of the two well-known
generalized-ensemble algorithms, namely,
multicanonical algorithm (MUCA)
and replica-exchange method (REM).
We then introduced four new generalized-ensemble
algorithms as further extensions of the above two methods,
which we refer to as replica-exchange multicanonical
algorithm (REMUCA), 
multicanonical replica-exchange method (MUCAREM),
multibaric-multithermal algorithm, and multi-overlap
algorithm.

With these new methods available,
we believe that we now have working simulation algorithms
for spin systems and biomolecular systems.


\noindent
{\bf Acknowledgments} \\
This work is supported, in part, by 
NAREGI Nanoscience Project, Ministry of Education, Culture, Sports,
Science and Technology, Japan.


\bibliographystyle{aipproc}   


\begin{thebibliography}{00}
\bibitem{RevHO} U.H.E. Hansmann and Y. Okamoto,
in {\it Annual Reviews of Computational Physics VI},
D. Stauffer (Ed.)
(World Scientific, Singapore, 1999) pp. 129--157.
\bibitem{RevMSO} A. Mitsutake, Y. Sugita, and Y. Okamoto,
{\it Biopolymers (Peptide Science)} {\bf 60}, 96--123 (2001).
\bibitem{RevSO} Y. Sugita and Y. Okamoto,
in {\it Lecture Notes in Computational Science and Engineering},
T. Schlick and H.H Gan (Eds.)
(Springer-Verlag, Berlin, 2002) pp. 304--332; cond-mat/0102296.

\bibitem{FS1} A.M. Ferrenberg and R.H. Swendsen,
{\it Phys. Rev. Lett.} {\bf 61}, 2635--2638 (1988); {\it ibid.}
{\bf 63}, 1658 (1989).
\bibitem{FSWHAM} A.M. Ferrenberg and R.H. Swendsen,
{\it Phys. Rev. Lett.} {\bf 63}, 1195--1198 (1989);
S. Kumar, D. Bouzida, R.H. Swendsen, P.A.
Kollman, J.M. Rosenberg, {\it J. Comput. Chem.}
  {\bf 13}, 1011--1021 (1992).

\bibitem{MUCA} B.A. Berg and T. Neuhaus, {\it Phys. Lett.} {\bf B267},
249--253 (1991);
{\it Phys. Rev. Lett.} {\bf 68}, 9--12 (1992).
    
\bibitem{Lee} J. Lee, {\it Phys. Rev. Lett.} {\bf 71}, 211--214
(1993); {\it ibid.} {\bf 71}, 2353.
\bibitem{MZ} M. Mezei, 
 {\it J. Comput. Phys.} {\bf 68}, 237--248 (1987).
\bibitem{BK} C. Bartels and M. Karplus,
 {\it J. Phys. Chem. B} {\bf 102}, 865--880 (1998).
 
\bibitem{Landau} F. Wang and D.P. Landau,
{\it Phys. Rev. Lett.} {\bf 86}, 2050--2053 (2001).
   
\bibitem{dePablo} N. Rathore and J.J. de Pablo,
{\it J. Chem. Phys.} {\bf 116}, 7225--7230 (2002);
Q. Yan, R. Faller, and J.J. de Pablo,
{\it J. Chem. Phys.} {\bf 116}, 8745--8749 (2002).

\bibitem{HO} U.H.E. Hansmann and Y. Okamoto,
{\it J. Comput. Chem.} {\bf 14}, 1333--1338 (1993).

5{\it J. Chem. Phys.}
5{\bf 103} (1995) 10298.


\bibitem{RE1} K. Hukushima and K. Nemoto,
{\it J. Phys. Soc. Jpn.} {\bf 65}, 1604--1608 (1996);
K. Hukushima, H. Takayama, and K. Nemoto,
{\it Int. J. Mod. Phys. C} {\bf 7}, 337--344 (1996).
   
\bibitem{RE2} C.J. Geyer, C.J. in {\it Computing Science and Statistics:
 Proc. 23rd Symp. on the Interface}, E.M. Keramidas (Ed.)
 (Interface Foundation, Fairfax Station, 1991) pp. 156--163.
\bibitem{RE3} R.H. Swendsen and J.-S. Wang, (1986)
 {\it Phys. Rev. Lett.} {\bf 57}, 2607--2609 (1986).
\bibitem{KT} K. Kimura and K. Taki, 
in {\it Proc. 13th IMACS World Cong. on Computation and Appl. Math. (IMACS '91)}, R. Vichnevetsky and J.J.H. Miller (Eds.),
vol. 2, pp. 827--828 (1991).
\bibitem{JWK} D.D. Frantz, D.L. Freeman, and J.D. Doll, 
{\it J. Chem. Phys.} {\bf 93}, 2769--2784 (1990).

\bibitem{Whit} M.C. Tesi, E.J.J. van Rensburg, E. Orlandini,
S.G. Whittington, {\it J. Stat. Phys.} {\bf 82}, 155--181 (1996).

\bibitem{STrev} E. Marinari, G. Parisi, J.J.
Ruiz-Lorenzo,
in {\it Spin Glasses and Random Fields}, A.P. Young
(Ed.) (World Scientific, Singapore, 1998) pp. 59--98.

\bibitem{IBArev} Y. Iba, {\it Int. J. Mod. Phys. C} {\bf 12},
623--656 (2001).

\bibitem{H97} U.H.E. Hansmann, 
{\it Chem. Phys. Lett.} {\bf 281}, 140--150 (1997). 
  
\bibitem{SO} Y. Sugita and Y. Okamoto, 
{\it Chem. Phys. Lett.} {\bf 314}, 141--151 (1999).
   
\bibitem{IRB2} A. Irb{\"a}ck and E. Sandelin, {\it J. Chem. Phys.}
{\bf 110}, 12256--12262 (1999).
\bibitem{FD} M.G. Wu and M.W. Deem, 
{\it J. Chem. Phys.} {\bf 111}, 6625--6632 (1999). 
\bibitem{Kol} D. Gront, A. Kolinski, and J. Skolnick, 
{\it J. Chem. Phys.} {\bf 113}, 5065--5071 (2000).
\bibitem{Gar} A.E. Garcia and K.Y. Sanbonmatsu, 
{\it Proteins} {\bf 42}, 345--354 (2001).
\bibitem{Berne} R.H. Zhou and B.J. Berne,
{\it Proc. Natl. Acad. Sci. U.S.A.} {\bf 99}, 12777--12782 (2002).

\bibitem{SKO} Y. Sugita, A. Kitao, and Y. Okamoto,
{\it J. Chem. Phys.} {\bf 113}, 6042--6051 (2000).

\bibitem{SO3} Y. Sugita and Y. Okamoto,
{\it Chem. Phys. Lett.} {\bf 329}, 261--270 (2000).
\bibitem{MO4} A. Mitsutake and Y. Okamoto,
{\it Chem. Phys. Lett.} {\bf 332}, 131--138 (2000).
\bibitem{MSO03} A. Mitsutake, Y. Sugita, and Y. Okamoto,
{\it J. Chem. Phys.} {\bf 118}, 6664--6675 (2003);
{\it ibid.} {\bf 118}, 6676--6688 (2003).
\bibitem{OO03} H. Okumura and Y. Okamoto,
submitted for publication; cond-mat/0306144.
\bibitem{BNO03} B.A. Berg, H. Noguchi, and Y. Okamoto,
{\it Phys. Rev. E}, in press; cond-mat/0305055.

\bibitem{Metro} N. Metropolis, A.W. Rosenbluth, M.N., Rosenbluth, 
A.H. Teller,  Teller, E. (1953) {\it J. Chem. Phys.} {\bf 21},
1087--1092.
   
\bibitem{mcd72}
I.R. McDonald, 
{\it Mol. Phys.} {\bf 23}, 41 (1972).
   
\bibitem{HMO97} U.H.E. Hansmann, M. Masuya, and Y. Okamoto,
{\it Proc. Natl. Acad. Sci. U.S.A.} {\bf 97}, 10652--10656 (1997).
   
\bibitem{NSMO} T. Nagasima, Y. Sugita, A. Mitsutake, and Y. Okamoto,
in preparation.

\bibitem{Bax} R.J. Baxter, {\it J. Phys. C} {\bf 6}, L445 (1973).

\bibitem{MUCAW} B.A. Berg,
{\it Nucl. Phys. B} (Proc. Suppl.) {\bf 63A-C}, 982--984 (1998).

\bibitem{jzg93}
J.K. Johnson, J.A. Zollweg, and K.E. Gubbins,
{\it Mol. Phys.} {\bf 78}, 591 (1993). 
%
\bibitem{sj93}
T. Sun and A.S. Teja,
{\it J. Phys. Chem.} {\bf 100}, 17365 (1996). 
\bibitem{OKK92} Y. Okamoto, K. Kikuchi, and H. Kawai,
{\it Chem. Lett.} {\bf 1992}, 1275--1278 (1992);
\bibitem{MHO98} A. Mitsutake, U.H.E. Hansmann, and Y. Okamoto, 
{\it J. Mol. Graphics Mod.} {\bf 16}, 226--238; 262--263 (1998);
   
\bibitem{ECEP3} 
M.J. Sippl, G. N{\'e}methy, and H.A. Scheraga,   
{\it J. Phys. Chem.} {\bf 88}, 6231--6233 (1984), and
references therein.
     
\bibitem{HOO99} U.H.E. Hansmann, Y. Okamoto, and J.N. Onuchic,
{\it Proteins} {\bf 34}, 472--483 (1999).
  
\bibitem{RasMol} 
R.A. Sayle and E.J. Milner-White,
{\it Trends Biochem. Sci.} {\bf 20}, 374--376 (1995).

\end{thebibliography}


\end{document}